\documentclass[journal,twoside,web]{ieeecolor}
\usepackage{generic}
\usepackage{cite}
\usepackage{amsmath,amssymb,amsfonts}
\usepackage{algorithmic}
\usepackage{graphicx}
\usepackage{algorithm,algorithmic}
\usepackage{hyperref}
\hypersetup{hidelinks=true}
\usepackage{caption}
\usepackage{subcaption}
\usepackage{textcomp}
\usepackage{diagbox}
\usepackage{siunitx}
\def\BibTeX{{\rm B\kern-.05em{\sc i\kern-.025em b}\kern-.08em
		T\kern-.1667em\lower.7ex\hbox{E}\kern-.125emX}}
\usepackage{mathtools}
\usepackage{booktabs}
\usepackage{multirow}

\bstctlcite{IEEEexample:BSTcontrol}

\def\x{\boldsymbol{x}}
\def\h{\boldsymbol{h}}
\def\z{\boldsymbol{z}}
\def\g{\boldsymbol{g}}
\def\p{\boldsymbol{p}}

\DeclarePairedDelimiter\parentheses{\lparen}{\rparen}

\newcommand{\cossit}[4]{\operatorname{s}_{{\tau \parentheses*{#1,#2}}}^{\parentheses*{#3,#4}}}

\newcommand{\encoder}[1]{\operatorname{H}_{\boldsymbol{\theta}}\parentheses*{#1}}

\begin{document}

\title{A comprehensive evaluation of pretraining strategies for channel-agnostic contrastive self-supervision of biosignals} 
\author{Thea Brüsch, Mikkel N. Schmidt, Tommy S. Alstrøm
\thanks{This paragraph of the first footnote will contain the date on which you submitted your paper for review. It will also contain support information, including sponsor and financial support acknowledgment. This work was supported by the Department of Applied Mathematics and Computer Science at the Technical University of Denmark.}
\thanks{Thea Brüsch, Mikkel N. Schmidt, and Tommy S. Alstrøm are with the Department of Applied Mathematics and Computer Science, Technical University of Denmark, Richard Petersens Plads 324, 2800 Kgs. Lyngby, Denmark (email: theb@dtu.dk, mnsc@dtu.dk, tsal@dtu.dk).}
\thanks{This paragraph will include the Associate Editor who handled your paper.}}


\maketitle

\begin{abstract}
Contrastive learning yields impressive results for self-supervision in computer vision. The approach relies on the creation of positive pairs, something which is often achieved through augmentations. However, for multivariate time series effective augmentations can be difficult to design. Additionally, the number of input channels for biosignal datasets often varies from application to application, limiting the usefulness of large self-supervised models trained with specific channel configurations. 
Motivated by these challenges, we set out to investigate strategies for creation of positive pairs for channel-agnostic self-supervision of biosignals. 
We introduce contrastive random lead coding (CRLC), where random subsets of the input channels are used to create positive pairs and compare with using augmentations and neighboring segments in time as positive pairs. We validate our approach by pre-training models on EEG and ECG data, and then fine-tuning them for downstream tasks. CRLC outperforms competing strategies in both scenarios in the channel-agnostic setting. Notably, for EEG tasks CRLC surpasses the current state-of-the-art reference model.  While, the state-of-the-art reference model is superior in the ECG task, incorporating CRLC allows us to obtain comparable results. In conclusion, CRLC helps generalization across variable channel setups when training our channel-agnostic model. The code is available at \url{https://github.com/theabrusch/Multiview_TS_SSL}.
\end{abstract}


\section{Introduction}
\label{sec:intro}

In recent times, self-supervised learning has demonstrated significant potential in the realms of computer vision and natural language processing~\cite{simclr, bert}. Self-supervised learning relies on inherent data patterns for pretraining on extensive, unlabeled datasets, facilitating the transfer of acquired knowledge to smaller labeled datasets, commonly referred to as downstream tasks. The acquisition of ground truth scoring for biosignals, such as electroencephalography (EEG) or electrocardiography (ECG), often demands the expertise of multiple professionals, making label acquisition a formidable and costly undertaking~\cite{younes2017a}. Consequently, self-supervised learning methods hold particular appeal for handling biomedical time series data.

An often used method for self-supervised learning is \emph{contrastive learning}. In contrastive learning, the model is presented with \emph{positive pairs} and \emph{negative pairs} and optimized such that positive pairs are close in representation space while negative pairs should be distant~\cite{1640964}. Other methods for self-supervised learning use reconstruction of masked inputs \cite{MAE} or require only positive pairs \cite{byol, barlowtwins}. While differing in optimization strategies, all methods for self-supervised learning share the same goal of learning generalizable representations from unlabelled data. In this work, we focus on \emph{contrastive self-supervised learning}.  


The big question in most contrastive learning centers around the optimal creation of positive and negative pairs, respectively. For images, the positive pairs may be created using augmentations like cropping, resizing, rotation, noise addition, and similar \cite{simclr}. Additional strategies involve obtaining multiple camera angles of the same object and using the different views to create positive pairs \cite{cameramultiview}. However, for time series data the design of augmentation strategies is less trivial and often requires specific domain knowledge.

Previously, a variety of methods have been proposed for constructing positive pairs from time series data. Some methods employ augmentations like masking, scaling, or random additive noise. Another strategy relies on contrastive predictive coding (CPC) using an autoregressive model to predict future sub-samples in time. A closely aligned tactic combines masking with CPC to reconstruct obscured segments within the sequence. 
Finally, some time series datasets contain different modalities or multiple channels, which can be interpreted as multiple views and thus be used to construct positive pairs. 

One of the main hurdles for self-supervised learning in multivariate biomedical time series is that the number of channels will often vary between different applications. 
This inconsistency poses challenges in transferring knowledge between tasks characterized by distinct channel configurations~\cite{9492125}, and existing methodologies often lack systematic approaches to address this concern. Previous work either discards excess channels or zero-pads missing channels during fine-tuning and/or pretraining~\cite{zhang2022selfsupervised, BENDR, leadagnosticecg}. Other methods pretrain their models on individual channels and only learn to combine channels in the downstream task ~\cite{seqclr, kiyasseh2021clocs}. 
In a recently published article \cite{10285993}, we demonstrate the promise of using a message passing neural network (MPNN) to extract inter-channel information to use both during pretraining and fine-tuning. In the original article, we only consider construction of positive pairs by forming groups of different channels. This work can be seen as an extension of our original paper. In this work, we use the proposed architecture to do a thorough investigation on the optimal way of forming positive pairs for multivariate time series data and aim to uncover under which assumptions different strategies hold. Our contributions include:

\begin{itemize}
    \item The introduction of a new synthetic dataset for testing methods for contrastive self-supervised learning on multivariate time series data. 
    \item A thorough investigation of optimal formation of positive pairs for pretraining and subsequent fine-tuning on multivariate time series data.
    \item Demonstration of the usability of the channel-agnostic architecture and the superiority of contrastive random lead coding for positive pair generation in two biomedical time series domains, namely EEG and ECG. 
\end{itemize}

\section{Related work}
Self-supervised learning for time series has gained more attention over the past few years. Within contrastive pretraining, previous significant work includes Eldele et al.~\cite{ijcai2021-324} who applied augmentations such as permutations and scaling to create positive pairs. They additionally used a temporal contrastive strategy resembling CPC to predict future augmented samples. In Yue et al. \cite{TS2Vec} random cropping and masking was used in combination with a new hierarchical time series loss to train their TS2Vec model. 
Deldari et al.~\cite{cocoa} leveraged the multi-view strategy for creating positive pairs by using different sensor modalities as positive pairs.
Banville et al. \cite{Banville_2021} compared different pretraining strategies for EEG data, but only investigated the performance when pretraining and fine-tuning on the same datasets, thus alleviating the channel mismatch issues faced when transferring between data settings. This is the same case for \cite{TS2Vec} and \cite{cocoa}, and while Eldele et al. and Kiyasseh et al. did investigate cross-dataset transfer, it was only shown for single-channel data. 

Zhang et al. \cite{zhang2022selfsupervised} applied similar augmentations to \cite{ijcai2021-324}, but additionally added a frequency encoder with accompanying augmentations. Additionally, they used a contrastive loss to enforce time-frequency consistency. 
BErt-like Neurophysiological Data Representation (BENDR) by Kostas et al.~\cite{BENDR} is inspired by self-supervised methods for natural language processing. The model consists of a convolutional encoder that takes in raw input EEG, which is subsequently processed by a transformer. The pretraining strategy is a combination of CPC and masking. 
While both of these methods demonstrate the ability to transfer across datasets with varying input channels, none of them are specifically tailored to do so. In both models, pretraining was done on a fixed sized input and excess channels in downstream tasks are removed while missing channels are zero-padded. 

Kiyasseh et al. \cite{kiyasseh2021clocs} and Mohsenvand et al. \cite{seqclr} both handled cross-dataset discrepancies in channel combinations by training a single-channel encoder and concatenating representations in the fine-tuning phase. Kiyasseh et al. used different channels and neighboring samples in time to create positive pairs in ECG, while Mohsenvand et al. applied augmentations to create positive EEG pairs. 

Finally, Oh et al. \cite{leadagnosticecg} used random lead masking on ECG signals during pretraining to prepare the model for fewer channels during fine-tuning. While the random lead masking helps the model to generalize when presented with fewer channels during fine-tuning, the model still takes a fixed number of input channels, limiting its usability in settings where there are more channels available for fine-tuning.   

This paper is a continuation of our previous work, where we used a message passing neural network (MPNN) to mitigate the channel-mismatch \cite{10285993}. Additionally, the flexibility of the MPNN allowed us to create positive pairs for contrastive learning using randomly sampled subsets of the input channels. In this paper, we extend the work by comparing different pretraining strategies in the variable-channel setting and test the framework on more datasets. 

\begin{figure*}
    \centering
    \includegraphics[width=\textwidth]{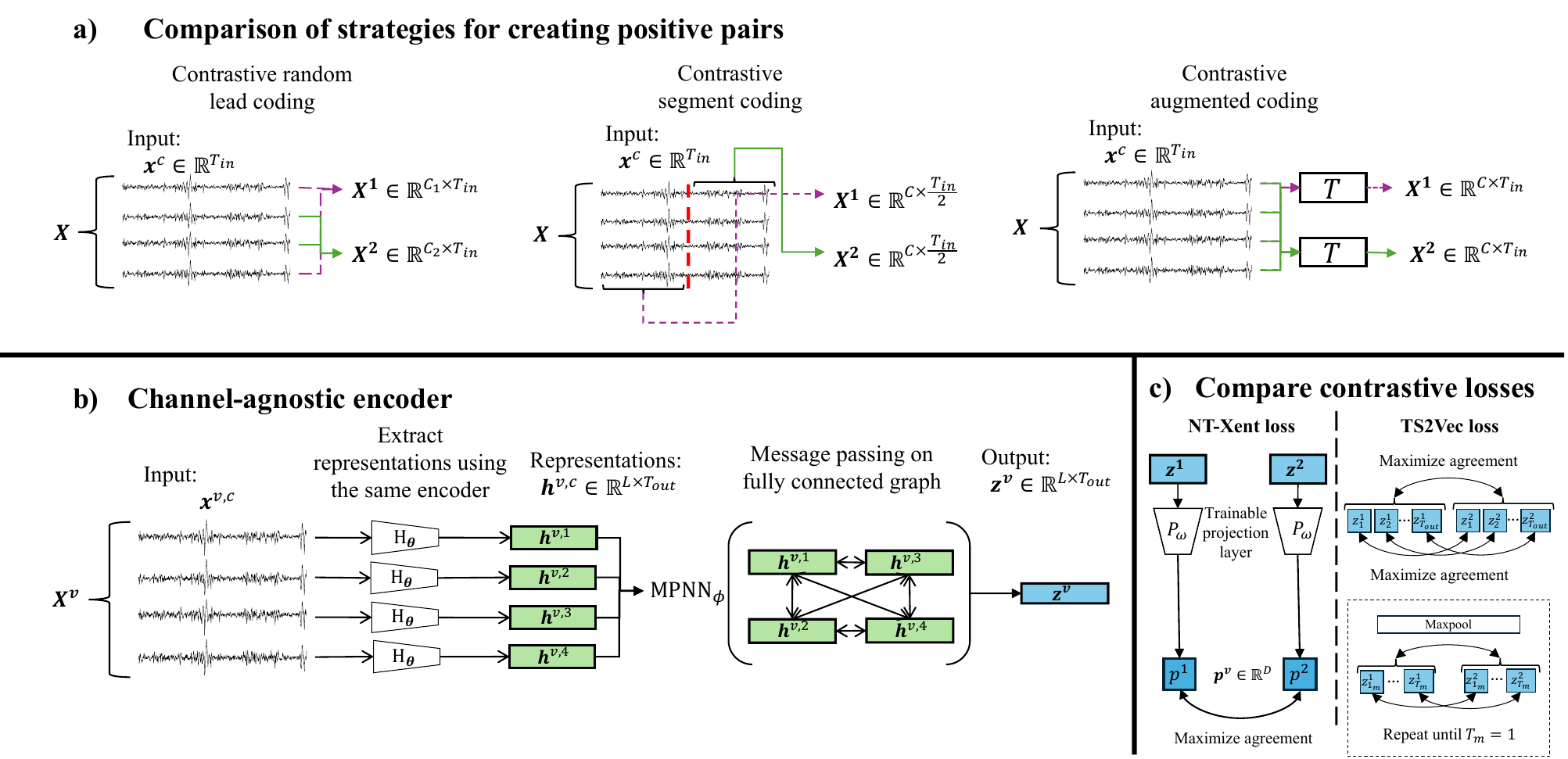}
    \caption{Overview of the experimental setup in this paper. a) We investigate three different strategies for creating positive pairs for multivariate biomedical time series, contrastive random lead coding (CRLC), contrastive segment coding (CSC) and contrastive augmented coding (CAC). b) We use a channel-agnostic neural network to extract representations that allow for generalization across variable channel setups. c) We compare two different contrastive losses for training the model in a self-supervised manner.}
    \label{fig:figure1}
\end{figure*}
\section{Methods}
In this section, we introduce the methods used throughout this paper to investigate optimal contrastive learning for multivariate biomedical time series. 

The section is organized as follows: First we present the channel-agnostic model architecture, slightly adapted from the model originally presented in \cite{10285993}, which allows us to generalize across variable channel setups. Second, we introduce the three strategies that we use to create positive pairs for contrastive learning, namely contrastive random lead coding (CRLC), contrastive segment coding (CSC), and contrastive augmented coding (CAC). Finally, we present the two contrastive losses which we use to train the models, the NT-Xent loss and the TS2Vec loss. An overview of the presented methods is shown in \autoref{fig:figure1}.
\subsection{Channel-agnostic model architecture}
We use a convolutional encoder to extract representations from the raw signals. We use the same convolutional encoder as in our original paper \cite{10285993} - an architecture inspired by the BENDR encoder \cite{BENDR}. For an input $\boldsymbol{X} \in \mathbb{R}^{C \times T_{\text{in}}}$ with $C$ channels, and length $T_{\text{in}}$, the same encoder, $\operatorname{H}_{\boldsymbol{\theta}}$, is used for each $\boldsymbol{x}^c$ to obtain the representation $\h^{c}$:
\begin{equation}
    \h^{c} = \encoder{\x^{c}}, \quad  \x^{c} \in \mathbb{R}^{T_{\text{in}}}, \quad \h^{c} \in \mathbb{R}^{L \times T_{\text{\text{out}}}},
\end{equation}
with output dimension, $L$, and a downsampled output length of $T_{\text{out}}$.

After the convolutional encoder, we use a message passing neural network (MPNN) to extract the inter-channel information. 
MPNNs were originally formalized in \cite{mpnn} and we follow their definition. An MPNN acts on graphs, and in our case, we form a fully connected undirected graph, $\g$, from the encoded representations of each channel in the input. This means that the graph consists of $C$ vertices with vertex features $\h$. 

The MPNN consists of two phases: the message passing phase, and the readout phase. The message passing is done over $K$ rounds, through message passing networks $\operatorname{M}_{\boldsymbol{\phi}_{k}}$ and update operations $\operatorname{U}_{k}$.
In each round, the message is obtained and used to update the state across all $\h$ in $\g$:
\begin{equation}
\begin{aligned}
\boldsymbol{m}_{k+1}^{\h} & =\frac{1}{C-1}\quad\sum_{\mathclap{\h'\in \g\backslash \{\h \}}} \operatorname{M}_{\boldsymbol{\phi}_{k}}\left(\h_{k}, \h_{k}' \right), \\
\h_{k+1} & =\operatorname{U}_{k}\parentheses*{\h_{k}, \boldsymbol{m}_{k+1}^{\h} }.
\end{aligned}
\end{equation}
We use a simple addition for the update operation, $\operatorname{U}_{k}\parentheses*{\h, \boldsymbol{m}^{\h}} = \h + \boldsymbol{m}^{\h}$. $\operatorname{M}_{\boldsymbol{\phi}_{k}}$ acts on the concatenation of $\h_{k}$ and $\h_{k}'$.
After the message passing, the readout network, $\operatorname{R}_{\boldsymbol{\phi}_{R}}$, computes the final output:
\begin{equation}
    \z = \operatorname{R}_{\boldsymbol{\phi}_{R}}\parentheses*{\frac{1}{C}\sum_{\h \in \g}\h_{K}}.
\end{equation}

In each round, $k$, the same network $\operatorname{M}_{\boldsymbol{\phi}_{k}}$ is used to pass messages between all nodes. Additionally, the readout network $\operatorname{R}_{\boldsymbol{\phi}_{R}}$ produces the final output from an average across all final hidden states. These two qualities combined make the $\operatorname{MPNN}_{\boldsymbol{\phi}}$ invariant to the number of input nodes and allows it to extract intra-graph representations from arbitrarily sized graphs. 
Therefore, the encoder is effectively channel-agnostic, enabling pretraining and fine-tuning on datasets with variable sets of input channels, without disregarding the inter-channel information in the pretraining phase. 

We refer to the full encoder, comprising $\operatorname{H}_{\boldsymbol{\theta}}$ and $\operatorname{MPNN}_{\boldsymbol{\phi}}$, as $\operatorname{G}_{\boldsymbol{\psi}}$:
\begin{equation}
 \operatorname{G}_{\boldsymbol{\psi}}\left(\boldsymbol{X}\right) = \operatorname{MPNN}_{\boldsymbol{\phi}}\left(\{\operatorname{H}_{\boldsymbol{\theta}}\left(\boldsymbol{x}^1\right), \dots, \operatorname{H}_{\boldsymbol{\theta}}\left(\boldsymbol{x}^C\right) \} \right).
\end{equation}

The full channel-agnostic network is shown in \autoref{fig:figure1} for $C=4$ channels.

\subsection{Positive pair generation}
We investigate the best ways to form positive pairs for multivariate time series data under the suggested model. Positive pairs are a necessity for contrastive self-supervised learning and is a way to impose invariance into the model. Specifically, for a time window $\boldsymbol{X}_i \in \mathbb{R}^{C\times T_{\text{in}}}$, we are looking for two views $\boldsymbol{X}_i^v, v\in\{1,2\}$ to form a positive pair. 
We study three methods for creating positive pairs during contrastive pretraining, namely contrastive random lead coding (CRLC), contrastive segment coding (CSC), and contrastive augmented coding (CAC). All three strategies are schematically shown in \autoref{fig:figure1}a).

\textbf{Contrastive random lead coding} (CRLC) uses different leads from within the same time window to create positive pairs. Given an input window $\boldsymbol{X}_i \in \mathbb{R}^{C\times T_{\text{in}}}$, we create two views by sampling one subset of the input leads and use the remaining input leads to create the other view, i.e. $\boldsymbol{X}^1_i \in \mathbb{R}^{C_1\times T_{\text{in}}}$ and $\boldsymbol{X}^2_i \in \mathbb{R}^{C_2\times T_{\text{in}}}$, where $C_1, C_2 \geq 2$. This strategy assumes that groups of channels recorded at the same time share the same content.

\textbf{Contrastive segment coding} (CSC) uses windows (segments) that are subsequent in time to create positive pairs. Given an input window $\boldsymbol{X}_{i} \in \mathbb{R}^{C\times T_{\text{in}}}$, we split the window into two equally sized segments along the time axis, such that $\boldsymbol{X}^1_i, \boldsymbol{X}^2_i \in \mathbb{R}^{C\times \frac{T_{\text{in}}}{2}}$. This strategy assumes that the content of the signal is stationary across the original window. 

\textbf{Contrastive augmented coding} (CAC) uses augmentations to create positive pairs. Specifically, we apply a series of transformations $\operatorname{T}=\operatorname{T_1}\circ\operatorname{T_2}\circ\dots\circ\operatorname{T_{N_T}}$ to create two augmented views of the input window $\boldsymbol{X}_{i}$, i.e. $\boldsymbol{X}^1_i = \operatorname{T}(\boldsymbol{X}_{i})$ and $\boldsymbol{X}^2_i = \operatorname{T}(\boldsymbol{X}_{i})$. All transformations contain random sampling of parameters, such that $\boldsymbol{X}^1_i \neq \boldsymbol{X}^2_i$. This strategy assumes that the data content is invariant to the transformations. 


\subsection{Contrastive losses}
For each of the two settings, we pretrain a neural network with two different contrastive losses. Given two views of the same data $\boldsymbol{X}_i^1$ and $\boldsymbol{X}_i^2$, we produce embeddings $\boldsymbol{z}_i^1$ and $\boldsymbol{z}_i^2$. 
\\
\textbf{NT-Xent loss~\cite{simclr}}: Prior to computing the NT-Xent loss, we project the embeddings to a lower dimensional space through a learnable projector, $\operatorname{P}_{\omega}(\boldsymbol{z}^v) = \boldsymbol{p}^v \in \mathbb{R}^{D}$. Given a batch of $N$ samples consisting of the two views $\boldsymbol{P}^v \in \mathbb{R}^{N\times D}$, we use the two views as positive pairs. The remaining $2N-2$ samples across both views are used as negative examples. Let 
\begin{equation}
    \cossit{i}{j}{w}{v} = \frac{\p_i^{w}\cdot\p_j^{v}}{\tau\|\p_i^{w}\|\cdot\|\p_j^{v}\|}
\end{equation} 
denote the $\tau$-scaled cosine similarity between $\p_i^{w}$ and $\p_j^{v}$. The loss for one positive pair then becomes:

\begin{equation}
\small
\ell^{(w,v)}_{i}=\ln \frac{\exp\parentheses*{\cossit{i}{i}{w}{v}}}
{\sum\limits_{j}^{N} \exp\parentheses*{\cossit{i}{j}{w}{v}} + \sum\limits_{j\neq i}^{N} \exp\parentheses*{\cossit{i}{j}{w}{w}}}.
\end{equation}

We compute $\ell^{(w,v)}$ for all samples of positive pairs in the batch and average over them:
\begin{equation}
\small
   \mathcal{L}_{\text{NT-Xent}}^{(w,v)} = -\frac{1}{N}\sum_{i}^N \ell_i^{(w,v)}.
\end{equation}
This operation is repeated for both views in each positive pair:
\begin{equation}
\small
    \mathcal{L}_{\text{NT-Xent}} = \frac{1}{2} 
    \left(\mathcal{L}^{(1,2)}_{\text{NT-Xent}} + \mathcal{L}^{(2,1)}_{\text{NT-Xent}}\right).
\end{equation}
\\
\textbf{TS2Vec loss~\cite{TS2Vec}}: The TS2Vec loss does not project the embeddings into a lower dimensional space, but instead acts directly on the two views, $\z^1_{i}$, $\z^2_{i}$. Additionally, it is designed to consider the temporal structure in the representations. 
It does this by splitting the loss into a temporal loss and an instance loss through different compositions of negative samples. The temporal loss, $lt^{w,v}_{(i,t)}$ uses time stamps from within the same sequence as negative examples. The instance loss, $li^{(w,v)}_{(i,t)}$, alternatively, uses the remaining instances in the batch at the same time $t$ as negative samples:
\begin{equation}
\small
{
\begin{aligned}
\ell t_{(i,t)}^{(w, v)}&=\ln \frac{\exp \left(\z_{i, t}^{w} \cdot \z_{i, t}^{v}\right)}{\sum\limits_{t^{\prime}}^{T_{\text{out}}}\exp \left(\z_{i, t}^{w} \cdot \z_{i, t^{\prime}}^{v}\right)+\sum\limits_{t^{\prime}\neq t}^{T_{\text{out}}}\exp \left(\z^{w}_{i, t} \cdot \z^{w}_{i, t^{\prime}}\right)}, \\
\ell i_{(i,t)}^{(w,v)}&=\ln \frac{\exp \left(\z_{i, t}^{w} \cdot \z_{i, t}^{v}\right)}{\sum\limits_{j}^N\left(\exp \z^{w}_{i, t} \cdot \z_{j, t}^{v}\right)+\sum\limits_{j\neq i}^N \exp \left(\z^{w}_{i, t} \cdot \z^{w}_{j, t}\right)}. \\
\end{aligned}}
\end{equation}
The two losses are then summed together over time and batch, creating the dual loss,
\begin{equation}
    \mathcal{L}^{(w,v)}_{\text{dual}}=-\frac{1}{2 N T} \sum\limits_{i}^{N} \sum\limits_t^{T_{\text{out}}}\left(\ell t_{(i,t)}^{(w, v)}+\ell i_{(i,t)}^{(w,v)}\right).
\end{equation}
A maxpool operation is used to iteratively downsample the representations along the temporal dimension, recomputing the dual loss at each iteration to create the hierarchical loss, $\mathcal{L}^{(w,v)}_{\text{TS2Vec}}$~\cite{TS2Vec}. The loss is averaged across both views:
\begin{equation}
\small
    \mathcal{L}_{\text{TS2Vec}} = \frac{1}{2}\left( \mathcal{L}_{\text{TS2Vec}}^{(1,2)}+\mathcal{L}_{\text{TS2Vec}}^{(2,1)}\right).
\end{equation}

\section{Evaluation on simulated data}
In order to test the assumed properties of different pretraining strategies, we generate synthetic multivariate time series data. The pretraining data is generated according to one of two different assumptions: 
\begin{itemize}
    \item All variables share some of the same information, but the information changes over time. 
    \item Not all variables share the same information, but the information is stationary over time.
\end{itemize}
The first setting is meant to show a case where it makes sense to use different channels as positive views, i.e. what we call the CRLC setting. The second setting is meant to show a case where successive samples in time can be used as positive pairs, i.e. suited well for what we call the CSC setting. 


\subsection{Data simulator}
We generate data with $C$ variables, $\boldsymbol{x}^c \in \mathbb{R}^{T_{\text{in}}}$, that are a mixture of $M$ sources $\boldsymbol{s}_m \in \mathbb{R}^{T_{\text{in}}}$. 
The collection of variables $\boldsymbol{X}_i \in \mathbb{R}^{C\times T_{\text{in}}}$ are mixed from the sources $\boldsymbol{S}_i \in \mathbb{R}^{M\times T_{\text{in}}}$ using a linear mixing matrix $\boldsymbol{A}_i \in \mathbb{R}^{C \times M}$:
\begin{equation}
    \boldsymbol{X}_{i} = \boldsymbol{A}_i\boldsymbol{S}_{i} + \boldsymbol{\epsilon}_i,
\end{equation}
where $\boldsymbol{\epsilon}_i \in \mathbb{R}^{C\times T_{\text{in}}}$ is drawn from a normal distribution, $\epsilon_{i,t}^c\sim \mathcal{N}(0, \sigma^2)$.

The sources are defined as sine functions with randomly generated frequencies, $\boldsymbol{f}_i \in \mathbb{R}^M$:
\begin{equation}
    \boldsymbol{s}_{i,m} = \sin\left(f_{i,m} t\right).
\end{equation}
We randomly sample $\boldsymbol{f}_i$ from a uniform distribution conditioned on the pretraining setting (see more in Section \ref{sec:simpre}), such that $\boldsymbol{S}_i = \begin{bmatrix}\boldsymbol{s}_{i,1} & \boldsymbol{s}_{i,2} &\dots& \boldsymbol{s}_{i,M}\end{bmatrix}^T$. 
Finally, $\boldsymbol{A}_i$ is randomly sampled also conditioned on the pretraining setting (see more in Section \ref{sec:simpre}), and normalised to sum to 1 across all variables for each source.  

\subsection{Simulating pretraining data} \label{sec:simpre}
\begin{figure}[tbp]
    \centering
    \includegraphics[width = \linewidth]{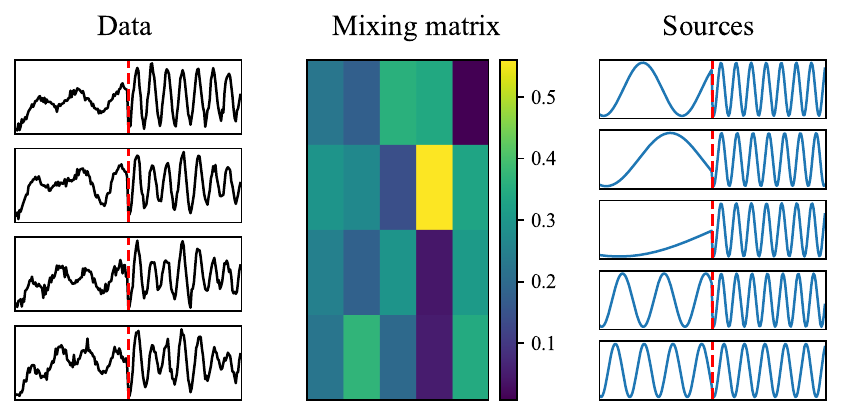}
    \caption{Example of simulated data designed to fit the contrastive random lead coding setting, but not the contrastive segment coding setting.}
    \label{fig:sim-CRLC}
\end{figure}

When simulating data that we hypothesize will fit well in the CRLC setting, we ensure that all output variables depend on most of the sources by randomly sampling the elements of $\boldsymbol{A}_i$ as $a^c_{i,m} \sim \mathcal{N}(0,1)$. We ensure that it does not fit with the CSC assumptions by sampling $\mathbf{f}_{i,t} \neq \mathbf{f}_{i, t+1}$, i.e. the state of the sources is changed between adjacent windows. These adjacent windows are only used when pretraining the models with the CSC loss. An example of such data sample is shown in \autoref{fig:sim-CRLC}.

\begin{figure}[tbp]
    \centering
    \includegraphics[width = \linewidth]{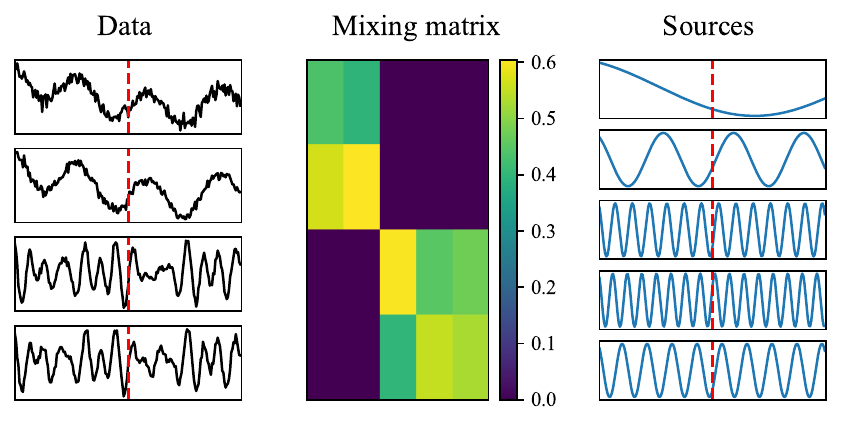}
    \caption{Example of simulated data designed to fit the contrastive segment coding setting, but not the contrastive random lead coding setting.}
    \label{fig:sim-csc}
\end{figure}

When simulating data that we hypothesize will fit well with the CSC setting, we ensure that adjacent windows share the same information by sampling $\mathbf{f}_{i,t} = \mathbf{f}_{i, t+1}$. Additionally, to ensure that the data does not follow the multiview assumptions, we make $\boldsymbol{A}_i$ block diagonal, thus creating two sets of output variables that each depend on two different sets of sources. An example of such data sample is shown in \autoref{fig:sim-csc}.

\subsection{Simulating fine-tuning data}
When generating data for fine-tuning, we follow a similar setup as for the pretraining dataset with a few exceptions. In this phase, we are mainly interesting in assessing the performance on two datasets: one that follows the CRLC hypothesis and one that does not. I.e. the main difference between the two settings is the mixing matrix which is either full, similar to the mixing matrix in \autoref{fig:sim-CRLC} or block diagonal as in \autoref{fig:sim-csc}. Additionally, the mixing matrix is sampled once prior to data generation and kept fixed for all samples coming from the same dataset. 

\begin{figure}[tbp]
    \centering
    \begin{subfigure}{\linewidth}
        \includegraphics[width=\linewidth]{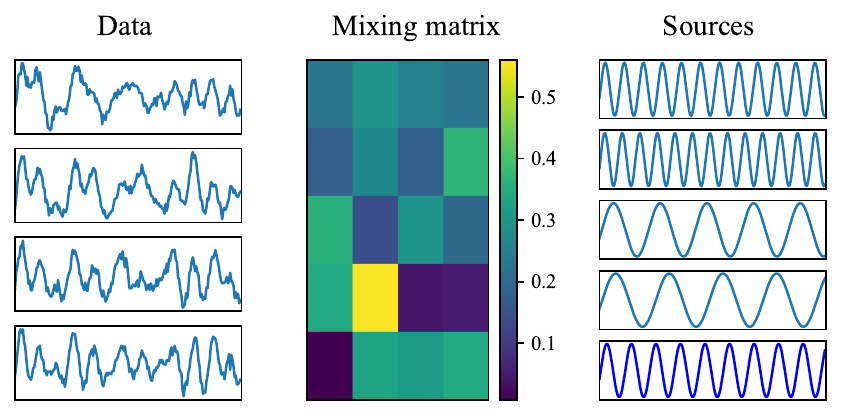}
    \caption{Class 1}
    \label{fig:enter-label}
    \end{subfigure}
    \begin{subfigure}{\linewidth}
        \includegraphics[width=\linewidth]{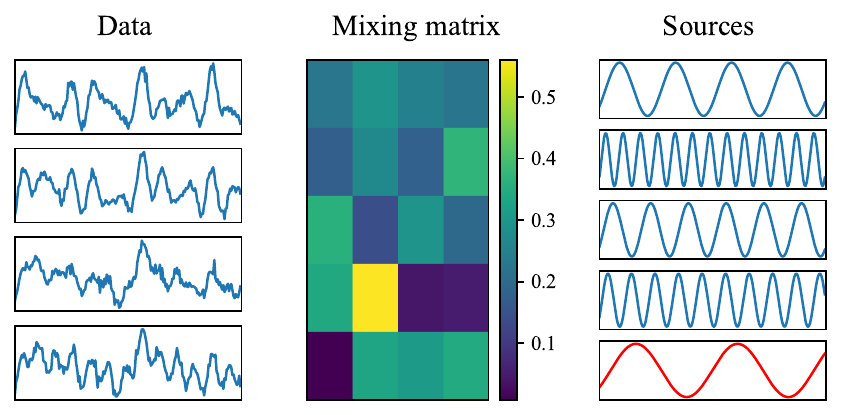}
    \caption{Class 2}
    \label{fig:enter-label}
    \end{subfigure}
    \caption{Simulated data for fine-tuning, where the class labels varies systematically based on the frequency of one source (the bottom source). The color of the source demonstrates the class.}
    \label{fig:sim-finetune}
\end{figure}

To generate labels for fine-tuning, we choose one source whose frequency varies systematically with the two classes. The frequencies of the remaining sources are chosen randomly for each data sample. An example of this is shown in \autoref{fig:sim-finetune}. Here, the bottom source determines the class, where the first class is marked with blue and the second class is marked with red.

\subsection{Experimental setup}
Both pretraining datasets are generated with 10 sources, $\boldsymbol{S}$, and 10 output variables, $\boldsymbol{X}$. For the CSC dataset, the mixing matrix is block diagonal with 5 sources and 5 output variables belonging to each group. We sample 10,000 windows of data for the pretraining dataset and 1000 for validation during pretraining. Each data window consists of 6000 samples in time and we use a variance of $\sigma^2=0.5^2$ when sampling the noise. 
We then train two models for each dataset: one using the CSC strategy to generate positive pairs and one using the CRLC strategy. When using the CRLC stategy to generate positive pairs, we only use the first 3000 samples in time during pretraining. This results in four pretrained models.
All models are pretrained for 50 epochs with a learning rate of $10^{-4}$ and a batch size of 32. 

These four pretrained models are tested on the two downstream datasets, one following the CRLC hypothesis and one that does not. We sample 10,000 samples for fine-tuning, 1,000 for validation, and 1,000 for testing. Again, we use 10 sources and 10 output variables and add noise with variance $\sigma^2=0.5^2$. The two different frequencies of the class-dictating source is chosen with $p=0.5$ to generate a class-balanced dataset. 

To evaluate the quality of the representations, we train a linear classifier on top of the learned representations while keeping the encoder frozen. We optimize the classifier with the AdamW optimizer with a learning rate of $10^{-3}$ and use early stopping on the validation accuracy to terminate training. 

\subsection{Results on simulated data}
The results on the downstream tasks are shown in \autoref{tab:synthresults}. Results are shown for fine-tuning on a dataset simulated with a block-diagonal mixing matrix as well as with a full mixing matrix. It is evident that pretraining with a strategy that does not fit the actual data generation process of the pretraining dataset yields poor representations. E.g. constructing positive pairs using the CSC strategy on a highly non-stationary CRLC pretraining dataset gives fine-tuning results that are only slightly better than random ($53.2-58.7\%$ accuracy across fine-tuning datasets.)

\begin{table}[tbp]     
\centering
\caption{Accuracy on the two downstream datasets, when fine-tuning either of the four pretrained models. The best results from each pretraining dataset are in bold.}
\begin{tabular}{llcccc}
\toprule
&&\multicolumn{4}{c}{Fine-tuning data}\\
&& \multicolumn{2}{c}{Block-diag. mix.} & \multicolumn{2}{c}{Full mixing}\\
\cmidrule(r){3-4}\cmidrule{5-6}
&& \multicolumn{2}{c}{Pre-train. data} & \multicolumn{2}{c}{Pre-train. data}\\
 && CSC & CRLC & CSC & CRLC\\
\midrule
\multirow{2}*{Pre-train. strategy} & CSC & 
$\boldsymbol{61.6\%}$  & 53.2\%& 57.2\%  & 58.7\% \\
& CRLC & 
54.3\%  & $\boldsymbol{78.0\%}$ & $\boldsymbol{58.5\%}$  & $\boldsymbol{74.8\%}$\\
\bottomrule
\end{tabular}

 \label{tab:synthresults}
\end{table}

The results for fine-tuning on data with block-diagonal mixing show that choosing the pretraining strategy that matches the downstream dataset increases performance on the downstream task. When fine-tuning on data constructed with a full mixing matrix, the results tell a slightly different story. With the CRLC pretraining on the CRLC dataset, the fine-tuned model reaches an accuracy of 74.8\%. However, the CSC pretraining on the CSC dataset only reaches an accuracy slightly above random and comparable with models trained with mismatch between data generation and pre-training strategy. A reason for this behaviour could be the standardization of the elements of the mixing matrix. The mixing matrix is standardized to sum to one for each source across all output variables. Thus when the contribution is spread across more output variables its contribution to each individual variable will be lower. This suggests that the mismatch between pretraining and fine-tuning dataset lowers the final performance. 

In real-world data one cannot design the pretraining dataset to fit well with the pretraining strategy; however, one can choose an appropriate pretraining strategy that fits well with the characteristics of the data used for pretraining. Additionally, our findings indicate that it is also important to consider the alignment between the pretraining and fine-tuning datasets.

\section{Evaluation on biomedical data}
Here, we evaluate the proposed setups across two different biomedical time series data types, electroencephalography (EEG) and electrocardiography (ECG).
\subsection{EEG data}
We pretrain on the Physionet Challenge 2018 (PC18) dataset \cite{yousnooze, PhysioNet}, originally annotated for sleep staging. The data set comprises six EEG channels (F3-M2, F4-M1, C3-M2, C4-M1, O1-M2, and O2-M1) sampled at 200 Hz. We resample everything to 100 Hz and divide the dataset into windows of 30s length without overlap. The subjects are split into training (80\%) and validation (20\%). In total we obtain 710,942 windows for pretraining and 178,569 windows for validation.

We then use two different datasets to fine-tune the model. The first dataset, SleepEDFx~\cite{PhysioNet, sleepedf}, comprises 153 nights of sleep recordings from 78 subjects with sleep stages manually annotated according by trained physicians. We combine N3 and N4 into one class, yielding five classes for classification (wake, N1, N2, N3 and R). The EEG data contains two channels (Fpz-Oz and Pz-Oz) with a sampling frequency of 100Hz. The data is scored based on 30s windows and thus we divide the entire dataset into windows of length 30s, again without overlap. We create a training, validation and test set by splitting the subjects into partitions of 0.6/0.2/0.2 and use the same splits throughout all experiments. This gives a total of 122,016 and 37,379 windows available for training and validation and 36,955 windows for testing. 

Additionally, we test the pretrained model by fine-tuning on the Montreal Archive of Sleep Studies (MASS) SS3 subset \cite{mass}\footnote{The dataset is open access but requires approval from the local ethics review board. The ethical approvement was obtained from the DTU Compute IRB Board with identicator number COMP-IRB-2024-01.}. 
This dataset consists of 62 whole-night EEG recordings from 62 different subjects. The EEG data contains the 20 channels adhering to the 10/20 montage. All channels are linked-ear referenced. We split the subjects 0.6/0.2/0.2 for training, validating, and testing and fix the splits. The windowing procedure yields 37,273 and 9,459 windows for training and validation, and 12,585 windows for testing.  
All windows in the pretraining and fine-tuning datasets are standardized, so each channel within each window has zero mean and a standard deviation of one. 

\subsection{ECG data}
\label{sec:ecgdata}
For ECG, we use the Physionet Challenge 2021 (PC21) database \cite{physio2021, willtwodo, PhysioNet}, which is a collection of 8 12-lead ECG datasets: The Chapman Shaoxing dataset, the CPSC  and the CPSC-Extra datasets, the PTB and PTB-XL datasets, the Georgia dataset, the Ningbo dataset, and the INCART dataset. We use the CPSC, CPSC-Extra, PTB-XL, Georgia, and Ningbo datasets for pretraining the model. We use the Chapman Shaoxing dataset for fine-tuning the model, i.e. this dataset is not included in the pretraining set. 

We split the records into windows of 10s duration with no overlap. All data is resampled to 500Hz. 
For the pretraining dataset, we split the windows 0.8/0.2 for training and validation, resulting in 64,845 windows for pretraining and 16,241 for validation.
For the fine-tuning dataset, we split the data 0.6/0.2/0.2 for training, validating, and testing and use the same splits throughout all experiments. This results in 6,375 windows for training, 2,124 for validation, and 2,129 for testing. 

All windows in the pretraining and fine-tuning datasets (both EEG and ECG datasets) are standardized, so each channel within each window has zero mean and a standard deviation of one.

\subsection{Model architecture}
We use the same architecture as in \cite{10285993} with the exception of the added projection layer when using the NT-Xent loss. The architecture of the convolutional encoder is closely inspired by~\cite{BENDR} and comprises 6 convolutional blocks with each a 1D convolution, dropout, group normalization, and a GELU activation function. In the first layer, the kernel width is set to 3 and in the remaining layers to 2 with the stride always equal to the kernel width. All intermediate layers use 256 kernels and the final output dimension is 64. A convolutional layer is added at the end with kernel width and stride 1. The trainable projection layer is a simple linear layer with an output dimension of $D=32$. The projector is discarded after pretraining.  

The message passing networks $\operatorname{M}_{\boldsymbol{\phi}_{k}}$ of the MPNN, all consist of a single linear layer, a dropout layer, and a ReLU activation function. We set $K=3$. We apply the same weights at all time steps $t\in T_{\text{out}}$. The message passing networks act on two nodes at a time, and the dimension of the weights are therefore $2\cdot64\times64$. The readout network $\operatorname{R}_{\boldsymbol{\phi}_{R}}$ has two linear layers with a dropout layer and a ReLU activation function in between. 

\subsection{Pretraining setup}
All of the models are pretrained for 20 epochs using the AdamW optimizer, a learning rate of $10^{-3}$, and weight decay set to $10^{-2}$.
The dropout rate is set to 10\% and we use a batch size of 64. 

For the CRLC and CAC pretraining strategies, we input the entire window to the encoder, i.e. $\boldsymbol{X}_i \in \mathbb{R}^{C\times T_{\text{in}}}$, where $T_{\text{in}}=3000$ for the EEG data, and $T_{\text{in}}=5000$ for the ECG data. 
This gives an output dimension $\h_i^{c}\in\mathbb{R}^{N\times L=64\times T_{\text{out}}=33}$ for the EEG sequences and $\h_i^{c}\in\mathbb{R}^{N\times L=64\times T_{\text{out}}=54}$ in the ECG sequences. For the CSC pretraining strategy, the windows are divided into to two, i.e. $T_{\text{in}}=1500$ for the EEG data, and $T_{\text{in}}=2500$ for the ECG data, resulting in $T_{\text{out}}=17$ for the EEG data and $T_{\text{out}}=27$ for the ECG data. 

\subsubsection{EEG augmentations}
We choose suitable augmentations from the literature for each data modality. For the EEG data, we apply the same augmentations as in \cite{seqclr} adjusted to fit the sampling frequencies and window size of our inputs. The augmentations are: 
\begin{itemize}
    \item Amplitude scaling, with scaling, $a\sim \mathcal{U}(0.5,2)$.
    \item Time shifting, with shift in number of samples $n\sim \mathcal{U}(-38,38)$.
    \item DC shift, with shift, $b\in\mathcal{U}(-10,10)$.
    \item Zero masking, with $m$ samples masked $m\sim\mathcal{U}(0,112)$ with mask starting at a random point, $t^{*}$.
    \item Additive Gaussian noise, $\epsilon\sim\mathcal{N}(0, \sigma^2)$, $\sigma \sim \mathcal{U}(0.01,0.2)$.
    \item Bandstop filter with $5Hz$ width centred around a frequency $f\sim\mathcal{U}(2.8, 41.3)$.
\end{itemize}
\subsubsection{ECG augmentations}
For the ECG data, we apply the same augmentations as applied for the SimCLR version in~\cite{leadagnosticecg}.
The augmentations are:
\begin{itemize}
    \item Random lead masking, where each lead is masked with zeros with probability $p=0.5$. 
    \item Powerline noise, where a sine curve of either 50 or 60 Hz (chosen with $p=0.5$) and phase, $\psi\sim\mathcal{U}(0, 2\pi)$, is scaled with $a\sim\mathcal{U}(0,0.5)$ and added to the signal. 
    \item Additive Gaussian noise, $\epsilon\sim\mathcal{N}(0, \sigma^2)$, $\sigma \sim \mathcal{U}(0.01,0.5)$.
    \item Baseline shift, where a random segment with a length of $\approx$ 20\% of the signal length is shifted. Each channel is shifted individually by $\pm b$, $b\in\mathcal{U}(0,0.25)$. The sign of the shift is chosen for each channel with probability $p=0.5$.  
    \item Baseline wander, where cosines of randomly chosen frequencies are added to the signal with amplitudes $A\sim \mathcal{U}(A_{min}, A_{max})$, where $A_{min} = 0$ and $A_{max} = 0.5$. The same cosines are added to all channels, but the amplitude is randomly adjusted for each channel by multiplying the cosine with $A_c = \mathcal{N}(1, 0.5)$. We add a total of three cosines. 
\end{itemize}



\subsection{Fine-tuning}
\label{sec:fine-tuning}
During fine-tuning, the pretrained $\operatorname{MPNN}_{\boldsymbol{\phi}}$ is used to obtain a single representation across all channels for all data modalities. The flexibility of the MPNN allows us to fine-tune on any number of input channels larger than 2. When the final representation is obtained, we use an average pooling along the time dimension to shorten the sequence to $T=4$ and flatten the representation. Finally, we use a linear layer with a softmax activation to classify each window. 

For all models, we choose the optimal fine-tuning learning rate on the validation sets by performing a grid search across $\lambda \in \{5\cdot 10^{-3}, 1\cdot 10^{-3}, 5\cdot 10^{-4}, 1\cdot 10^{-4}, 5\cdot 10^{-5}\}$. The models are optimized with the AdamW optimizer with a fixed weight decay of $10^{-2}$ and batchsize 32.

We wish to investigate settings for which few labels are avaiable for fine-tuning. Therefore, we sample balanced datasets from the available windows in each dataset. From each class, we sample 10, 20, 50, 100, 200, 500, and 1000 windows for fine-tuning of our pretrained models as well as models trained from scratch. The validation sets are sampled similarly such that they contain the same number of samples as the training sets. We fine-tune all models for maximum 100 epochs, using early stopping on the validation accuracy to terminate training after convergence. We investigate two settings: freezing the encoder during fine-tuning and optimizing the entire model during fine-tuning.


\subsection{State-of-the-art reference models}
To properly assess the effect of the pretraining strategy the architecture is kept fixed for all models. However, to relate the recorded performance to existing literature, we additionally choose one representative state-of-the-art reference for each domain. Both of these models are inspired by wav2vec 2.0 \cite{wav2vec} (henceforth referred to as W2V), which uses a combination of contrastive learning and masking to pretrain an encoder for single-channel audio data.

For EEG data, we choose BENDR \cite{BENDR}, since the model has been shown to generalize between different channel setups. We use the publically available weights from the model pretrained on the large Temple University Hospital EEG data corpus \cite{temple}. The model has thus been designed to take a fixed input of 20 channels. Additional channels during fine-tuning are discarded, while missing channels are zero-padded. 

For ECG data, we use the model by Oh et al. \cite{leadagnosticecg} that the authors refer to as W2V+CMSC+RLM. Here, the pretrained weights are not publically available. We therefore pretrain the model using their code on the same dataset on which our own models have been pretrained. 
The W2V+CMSC+RLM model uses the wav2vec 2.0 setup to extract local representations for each window of data through masking and reconstrucion of learned tokens. Additionally, the authors adds a contrastive loss across windows of data by generating positive pairs. The authors refer to their strategy for positive pair generation as CMSC, however, this strategy is identical to the one we refer to as CSC. Therefore, we also train a version of this network using the CRLC strategy, by adapting their original code. 
\section{Results}

\subsection{EEG results}
We run the fine-tuning experiments for five different seeds (i.e., both the data sampling and classifier initialization are reseeded five times) and report the averaged scores. 

\begin{figure}[t]
    \centering
    \includegraphics{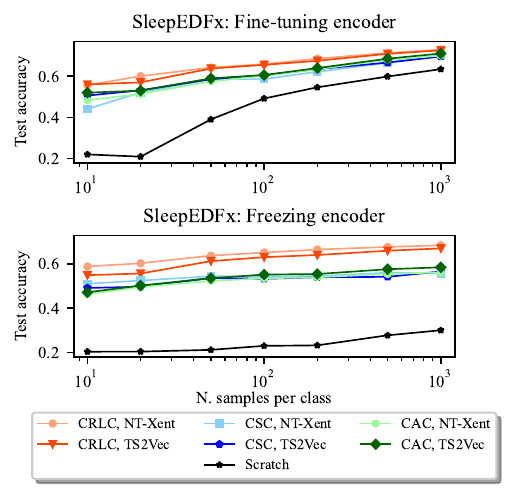}
    \caption{SleepEDFx results.}
    \label{fig:eeg_results}
\end{figure}

\autoref{fig:eeg_results} (top) shows the results on the SleepEDFx dataset when optimizing the entire network during fine-tuning. The results are also shown for a subset of samples in \autoref{tab:eeg_optenc} (left). All of the reported scores are balanced test accuracies on the fixed test set and chance level is 20\%.
Since all of the results, except for BENDR, are shown using the exact same network, we can directly compare the different strategies for positive pair generation. 
We see that all of the pretraining strategies improve the scores over the model trained from scratch. The improvement is most prominent for the low sample sizes, while the model trained from scratch closes the gap on the pretrained models once the sample size is increased. Additionally, we see that the BENDR model is outperformed by all of our pretrained models, indicating the benefit of the channel-agnostic setup. 

When comparing the two loss functions, we see that for the CSC and CAC pretraining strategies, the models trained using TS2Vec generally outperforms the models trained using the NT-Xent loss. The opposite is true for the model trained using the CRLC strategy, where the NT-Xent model actually outperforms the TS2Vec model at most sample sizes. However, the difference is not significant.  

Comparing pretraining strategies, it is clear from the results that the CRLC pretraining strategy outperforms the remaining two at all sample sizes. The CAC and CSC strategies perform comparably to each otehr with a slight preference towards the CAC strategy. Again, the biggest difference is observed at the smallest sample size, where the gap in percentage points from the best performing model to the worst performing pretrained model is 11.9\%. 

\begin{figure}[t]
    \centering
    \includegraphics{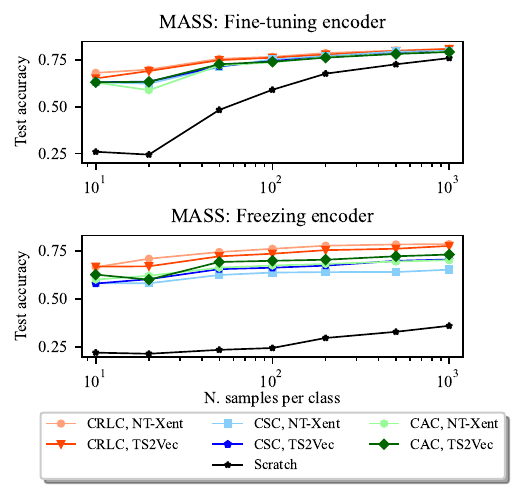}
    \caption{MASS results.}
    \label{fig:mass_results}
\end{figure}

A similar pattern emerges when assessing the performance on the MASS dataset in \autoref{tab:mass_optenc} (left) and \autoref{fig:mass_results} (top). Again, the CRLC strategy combined with the NT-Xent loss outperforms all other combinations listed in the table. Noticeably, the model again outperforms the BENDR reference model performance. Here, it is worth reiterating that our models are all pretrained on an EEG dataset with 6 channels, while the BENDR dataset is pretrained on an EEG dataset adhering to the 10/20 channels setup, i.e. 19 channels with a channel-specific encoder. The MASS dataset is similarly recorded using the 10/20 montage, but contains an additional channel (Oz). In theory, the BENDR model should be designed specifically for this setup, but fails to generalize as well as the channel-agnostic models. Additionally, this shows the unique capability of the MPNN that allows the model to be fine-tuned on more channels than what was present during pretraining without discarding additional information. 


\begin{table*}[t]
    \centering
    \footnotesize \caption{SleepEDFx: Balanced accuracy scores in \% after either optimizing the entire network during fine-tuning or freezing the encoder averaged across 5 seeds. The standard error of the mean is shown in parentheses. Results are shown for 4 different sample sizes, where the number of samples is chosen as the number of samples per class in both the training set and validation set. Highest accuracy for each sample size is marked in bold, while the second highest is underlined. The highest score is marked with a $^*$ if the results is signficantly ($p<0.05$) higher than the second best result.}
        \begin{tabular}{@{}llrrrrrrrrr@{}}
            \toprule
            Pretraining & Loss & \multicolumn{4}{c}{ Fine-tuning encoder (N samples per class)} & & \multicolumn{4}{c}{ Freezing encoder (N samples per class)}  \\
            \cmidrule(lr){3-6} \cmidrule(lr){8-11}
                           &  & N=10 & N=50 & N=100 & N=1000 & & N=10 & N=50 & N=100 & N=1000\\
            \midrule
            BENDR \cite{BENDR} & W2V & $24.2(1.8)$& $43.1(2.7)$& $49.5(2.6)$& $62.6(0.7)$ & & $20.3(0.1)$& $20.0(0.1)$& $20.3(0.1)$& $21.4(0.3)$ \\\hline
            None & None & $22.0(0.7)$& $39.1(1.4)$& $49.2(0.9)$& $63.5(0.5)$ &  & $20.3(0.2)$& $21.2(0.2)$& $23.0(1.0)$& $30.0(0.3)$ \\ 
            CRLC & NT-Xent & $\underline{55.7}(2.0)$& $\mathbf{64.3}(0.6)$& $\mathbf{65.9}(0.4)$& $\mathbf{73.0}(0.2)$ & & $^*\mathbf{59.0}(1.2)$& $ ^*\mathbf{63.9}(0.2)$& $^*\mathbf{65.3}(0.4)$& $^*\mathbf{68.6}(0.1)$ \\
            CRLC & TS2Vec & $\mathbf{56.0}(1.6)$& $\underline{63.8}(0.7)$& $\underline{65.6}(0.4)$& $\underline{72.6}(0.2)$ & & $\underline{55.0}(0.9)$& $\underline{61.4}(0.2)$& $\underline{63.2}(0.3)$& $\underline{67.1}(0.1)$ \\
            CSC & NT-Xent & $44.1(3.2)$& $58.6(0.5)$& $58.6(0.5)$& $70.3(0.5)$ &  & $51.1(0.5)$& $54.5(0.5)$& $54.4(0.6)$& $55.8(0.2)$ \\
            CSC & TS2Vec & $50.6(1.8)$& $58.6(0.7)$& $60.4(0.5)$& $69.6(0.3)$ & & $49.2(1.4)$& $53.7(0.6)$& $53.4(0.6)$& $56.7(0.6)$ \\
            CAC & NT-Xent & $48.4(2.3)$& $57.5(1.2)$& $60.3(0.6)$& $70.1(0.3)$& & $46.4(1.1)$& $52.4(0.1)$& $53.6(0.2)$& $56.3(0.2)$\\
            CAC & TS2Vec& $52.1(1.1)$& $59.0(0.7)$& $60.6(0.7)$& $71.1(0.5)$ & & $47.2(0.8)$& $53.7(0.3)$& $55.3(0.4)$& $58.6(0.3)$ \\
            \bottomrule
        \end{tabular}

    \label{tab:eeg_optenc}
\end{table*}

\begin{table*}[t]
    \centering
    \footnotesize
    \caption{MASS: Balanced accuracy scores in \% after either optimizing the entire network during fine-tuning or freezing the encoder averaged across 5 seeds. The standard error of the mean is shown in parentheses. Results are shown for 4 different sample sizes, where the number of samples is chosen as the number of samples per class in both the training set and validation set. Highest accuracy for each sample size is marked in bold, while the second highest is underlined. The highest score is marked with a $^*$ if the results is signficantly ($p<0.05$) higher than the second best result.}
        \begin{tabular}{@{}llrrrrrrrrr@{}}
            \toprule
            Pretraining & Loss & \multicolumn{4}{c}{ Fine-tuning encoder (N samples per class)} & & \multicolumn{4}{c}{ Freezing encoder (N samples per class)}  \\
            \cmidrule(lr){3-6} \cmidrule(lr){8-11}
                           &   & N=10 & N=50 & N=100 & N=1000 & & N=10 & N=50 & N=100 & N=1000\\
            \midrule
        BENDR \cite{BENDR} & W2V & $39.9(4.1)$& $60.5(1.2)$& $66.8(1.0)$& $78.4(0.2)$ & & $19.9(0.1)$& $21.0(0.4)$& $21.3(0.5)$& $22.1(1.0)$ \\ \hline
        None & None & $25.9(0.2)$& $48.3(3.1)$& $59.1(1.5)$& $76.0(0.3)$ & & $22.2(0.7)$& $23.6(0.9)$& $24.6(1.0)$& $36.1(0.7)$\\
        CRLC & NT-Xent & $\mathbf{68.1}(1.5)$& $\mathbf{75.6}(0.7)$& $\mathbf{76.7}(0.4)$& $\mathbf{81.1}(0.2)$ & & $\underline{66.6}(1.2)$& $^*\mathbf{74.4}(0.5)$& $^*\mathbf{76.1}(0.4)$& $^*\mathbf{78.5}(0.1)$ \\
        CRLC & TS2Vec & $\underline{65.1}(2.1)$& $\underline{74.9}(0.2)$& $\underline{76.3}(0.4)$& $\underline{80.9}(0.1)$ &  & $\mathbf{66.8}(0.7)$& $\underline{72.2}(0.4)$& $\underline{73.5}(0.4)$& $\underline{77.6}(0.3)$ \\
        CSC & NT-Xent & $63.1(2.3)$& $71.2(0.7)$& $75.4(0.5)$& $79.9(0.3)$ & & $58.3(0.9)$& $62.5(0.2)$& $63.8(0.1)$& $65.3(0.2)$ \\
        CSC & TS2Vec & $63.1(1.9)$& $71.6(0.6)$& $74.5(0.3)$& $79.6(0.1)$ & & $58.1(1.3)$& $65.5(0.4)$& $66.3(0.8)$& $70.6(0.5)$ \\
        CAC & NT-Xent & $63.0(1.7)$& $72.1(0.7)$& $73.7(0.4)$& $79.6(0.2)$ & & $60.1(0.8)$& $66.3(0.5)$& $67.4(0.2)$& $70.2(0.1)$ \\
        CAC & TS2Vec & $63.1(2.4)$& $72.7(0.7)$& $74.1(0.3)$& $79.3(0.3)$ & & $62.7(0.9)$& $69.3(0.1)$& $69.9(0.2)$& $73.1(0.2)$\\
        \bottomrule
    \end{tabular}
    
    \label{tab:mass_optenc}
\end{table*}

\autoref{fig:eeg_results} (bottom) shows the results on the SleepEDFx dataset when fine-tuning only the classifier layer of the model. Again, a subset of the results is shown in \autoref{tab:eeg_optenc} (right). Optimizing only the classifier allows us to assess the pure quality of the representations learned during pretraining. The representations are compared to random representations by using a randomly initialized encoder, which is also frozen during fine-tuning. Here, it is clear that the random representations are not very discriminative of the five sleep stages, as the model reaches a maximum accuracy of 30\%, which is barely higher than random guessing. Additionally, the BENDR model performs even worse than random representations in this setting. 

The pretrained models, however, again show that the pretraining extracts meaningful features for the downstream task. The best performing model is again the one trained with CRLC pretraining strategy with the NT-Xent loss. Comparing the numbers from \autoref{tab:eeg_optenc} (left) and \autoref{tab:eeg_optenc} (right) actually shows that for the smallest sample size, the fine-tuned model with the frozen encoder outperforms the fully fine-tuned model. Additionally, for most other sample sizes, the performance of the frozen encoder model is comparable to the non-frozen model. This indicates that the learned representations are useful as is, and that in some cases the frozen encoder actually helps regularize the model towards a better fit. Similar results are shown on the MASS dataset in \autoref{tab:mass_optenc} (right), though while the frozen encoder does not outperform the fully fine-tuned one, the results get very close. 

In summary, the results indicate that the CRLC method is superior for generating positive pairs for contrastive learning for EEG data. One could hypothesize that since different channels are used as positive views of each other, the model learns to disentangle the underlying signal that the signals have in common. Furthermore, the use of the MPNN makes the transition between variable channel setups seamless without the need for zero-padding or other alterations to the downstream dataset.

\subsection{ECG results}
We perform the same experiments for the ECG data. Similarly, here the results are averaged across 5 seeds and reported in terms of balanced accuracy scores on the fixed test set. 

\begin{figure}[t]
    \centering
    \includegraphics{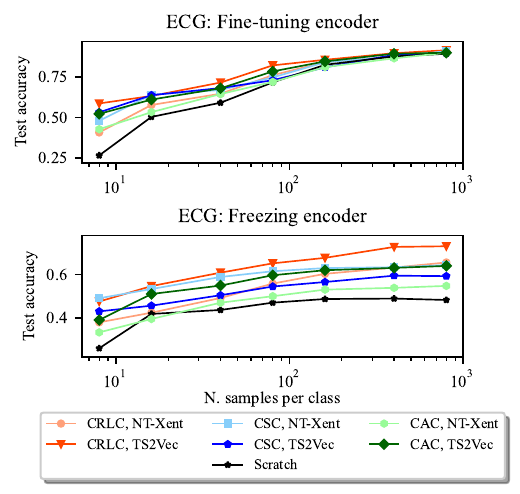}
    \caption{ECG results.}
    \label{fig:ecg_results}
\end{figure}

\autoref{fig:ecg_results} (top) and \autoref{tab:ecg_optimize_encoder} (left) show the results when optimizing the entire model during fine-tuning. 
Here it is clear that the state-of-the-art reference model outperforms all of the proposed pretraining strategies. These results show that the additional training of local representations within each sequence that happens through the masking and contextualization in W2V is a strong component in extracting ECG information. It is, however, also worth noticing that while the MPNN driven models contain $5.9\cdot10^4$ parameters, the W2V driven model contains $9.0\cdot10^7$ parameters. Additionally, if we compare the reference model adapted to the CRLC pretraining strategy, this model obtains the second highest or highest accuracy across fine-tuning strategies and sample sizes and is not significantly different from the original reference model. The model was adapted using minimal alterations and thus a more carefully designed scheme might yield better results. 

If we instead compare the pretraining strategies of the MPNN driven models, it is again clear that the CRLC strategy is superior to CSC and CAC. Here, this is only the case when combined with the TS2Vec loss. An explanation could be that since the TS2Vec loss aligns the positive pairs in a token-wise manner (instead of just on instance level), the pretraining task becomes more difficult and in turns makes the representations more useful. This loss can in some ways be compared to the W2V loss, which clearly is a powerful method for ECG pretraining. 

Again, we see that the difference in results is most prominent at small sample sizes, however the CRLC remains the strongest method across all sample sizes.

\begin{table*}[t]
    \centering
    \footnotesize
    \caption{ECG: Balanced accuracy scores in \% after either optimizing the entire network during fine-tuning or freezing the encoder averaged across 5 seeds. The standard error of the mean is shown in parentheses. Results are shown for 4 different sample sizes, where the number of samples is chosen as the number of samples per class in both the training set and validation set. Highest accuracy for each sample size is marked in bold, while the second highest is underlined. The highest score is marked with a $^*$ if the results is signficantly ($p<0.05$) higher than the second best result.}
    \begin{tabular}{@{}llrrrrrrrrr@{}}
            \toprule
            Pretraining &  Loss & \multicolumn{4}{c}{ Fine-tuning encoder (N samples per class)} & & \multicolumn{4}{c}{ Freezing encoder (N samples per class)}  \\
            \cmidrule(lr){3-6} \cmidrule(lr){8-11}
                           &   & N=10 & N=50 & N=100 & N=1000 & & N=10 & N=50 & N=100 & N=1000\\
       \midrule
       CSC \cite{leadagnosticecg} & W2V + NT-Xent & $\underline{81.5}(1.9)$ & $\mathbf{92.4}(0.5)$ & $\mathbf{94.1}(0.3)$ & $\mathbf{96.4}(0.2)$ & & $\mathbf{65.0}(1.9)$ & $\mathbf{84.2}(0.2)$ & $\mathbf{88.7}(0.4)$ & $\underline{94.9}(0.2)$\\ 
       CRLC & W2V + NT-Xent & $\mathbf{82.0}(0.8)$ & $\underline{91.6}(0.5)$ & $\underline{93.4}(0.9)$ & $\underline{96.2}(0.1)$ & & $\underline{60.0}(3.0)$ & $\underline{83.4}(0.9)$ & $\underline{88.0}(0.4)$ & $\mathbf{95.0}(0.1)$\\ \hline
       None & None & $26.5(0.8)$& $59.1(0.7)$& $71.6(2.1)$& $89.3(0.8)$ & & $25.9(0.7)$& $43.6(0.5)$& $47.0(1.2)$& $48.2(1.2)$ \\
        CRLC & NT-Xent & $40.7(4.7)$& $64.6(3.2)$& $76.1(0.3)$& $91.1(0.3)$ & & $37.9(1.5)$& $49.1(0.7)$& $55.8(0.4)$& $65.6(0.1)$ \\
        CRLC & TS2Vec & $\mathit{58.6}(0.8)$& $\mathit{71.5}(2.1)$& $\mathit{82.2}(0.5)$& $\mathit{91.5}(0.4)$ & & $47.5(2.5)$& $\mathit{60.8}(0.7)$& $\mathit{65.1}(0.4)$& $\mathit{73.0}(0.6)$ \\
        CSC & NT-Xent & $47.9(2.8)$& $66.9(1.4)$& $74.7(2.9)$& $91.0(0.4)$ & & $\mathit{49.0}(1.1)$& $58.9(0.8)$& $61.5(0.6)$& $64.3(0.3)$ \\
        CSC & TS2Vec & $53.2(1.8)$& $68.2(1.2)$& $73.0(1.0)$& $89.7(0.2)$ & & $43.0(1.4)$& $50.4(0.6)$& $54.4(0.5)$& $59.2(0.8)$ \\
        
        CAC & NT-Xent & $42.7(4.7)$& $64.4(1.3)$& $71.9(0.7)$& $89.9(0.2)$& & $33.3(2.2)$& $47.0(0.4)$& $50.0(0.8)$& $54.7(0.2)$\\
        CAC & TS2Vec & $52.2(1.4)$& $68.0(0.6)$& $78.4(1.2)$& $90.0(0.5)$ &  & $39.0(2.0)$& $54.9(0.4)$& $59.7(0.8)$& $64.0(0.5)$ \\
        \bottomrule
    \end{tabular}
    
    \label{tab:ecg_optimize_encoder}
\end{table*}

\autoref{fig:ecg_results} (bottom) and \autoref{tab:ecg_optimize_encoder} (right) show the results when freezing the encoder. Again, the reference models outperform all MPNN based models. 
Comparing only the MPNN models, the CSC strategy yields the highest accuracy on the smallest sample size, while the CRLC combined with the TS2Vec loss outperforms the other models on the remaining sample sizes. 

Overall, the results show that the larger state-of-the-art reference model outperforms the smaller models that have been used to compare strategies in this study. In addition to being larger, the reference model takes a fixed sized input, which allows the model to tailor the extracted representations of each individual channel in the dataset. In the reference model, channel-agnosticity is achieved by masking random leads during pretraining. This strategy allows for more seemless fine-tuning on downstream datasets with fewer input channels. However, this setup does not allow for fine-tuning on downstream datasets with more input channels, which is a downside compared to the MPNN based models. 

\subsection{Number of message passing rounds}
To examine the importance of aggregating information through message passing, we study the effect of the number of message passing rounds $K\in\{0,1,2,3\}$ before the mean aggregation. 
Results in \autoref{tab:ablation} show that for both EEG datasets, the method is not sensitive to the number of message passing rounds, yielding near-identical performance in all cases. For the ECG data the situation is similar, however showing a slight advantage of the message passing when the number samples per class for finetuning is 50 or more. These results indicate that the positive pair formation and contrastive loss have a greater impact on performance that the aggregation strategy.
\begin{table}[h]
    \centering
    \footnotesize
    \caption{Results with $K=3$ and $K=0$ message passing rounds. Best results for each dataset at each $N$ is marked in bold, while significant results are marked with $^*$.}
    \begin{tabular}{@{}lcrrrr@{}}
    \toprule
        Data & $K$ & \multicolumn{4}{c}{ Samples per class} \\
        \cmidrule(l){3-6}
                    &   &  10 & 50 & 100  & 1000 \\
       \midrule
       \multirow{4}*{EDF} & 0 & $56.6(1.5)$& $\mathbf{65.1}(0.4)$& $\mathbf{67.3}(0.2)$& $72.8(0.1)$  \\
       & 1 & $\mathbf{59.0}(1.6)$& $64.8(0.5)$& $67.0(0.6)$& $\mathbf{73.0}(0.3)$\\
       & 2 & $57.6(1.9)$& $64.3(0.9)$& $67.1(0.4)$& $72.8(0.3)$ \\
        & 3 & $55.7(2.0)$ & $64.3(0.6)$ & $65.9(0.4)$ & $\mathbf{73.0}(0.2)$ \\
       \midrule
        
       \multirow{4}*{MASS} & 0 &  $\mathbf{70.8}(1.0)$& $\mathbf{75.6}(0.4)$& $77.1(0.3)$& $81.4(0.1)$ \\  
       & 1 & $69.9(0.8)$& $75.1(0.3)$& $\mathbf{77.2}(0.3)$& $\mathbf{81.5}(0.1)$\\
       & 2 & $68.4(1.3)$& $74.3(0.6)$& $76.9(0.8)$& $81.3(0.2)$ \\
       & 3 & $68.1(1.5)$& $75.6(0.7)$& $76.7(0.4)$& $81.1(0.2)$\\
       \midrule
       \multirow{4}*{ECG} & 0 & $\mathbf{61.0}(0.8)$ & $70.8(2.1)$ & $77.4(0.8)$ & $89.2(0.1)$ \\
       
        & 1 & $58.3(1.6)$& $71.1(2.0)$& $81.8(0.7)$& $91.3(0.4)$ \\
        & 2 & $56.0(0.6)$& $\mathbf{74.3}(1.1)$& $82.0(0.5)$& $91.2(0.3)$ \\
       & 3 & $58.6(0.8)$ & $71.5(2.1)$  & $\mathbf{82.2}(0.5)$ & $\mathbf{91.5}(0.4)$  \\
        \bottomrule
    \end{tabular}
    
    \label{tab:ablation}
\end{table}


\section{Discussion and Future work}
Self-supervised learning for time series suffers from the high variability in datasets, even within the same domain. This is especially the case for EEG data, where channel setups may vary significantly between different settings. While big datasets, such as the TUH EEG corpus \cite{temple} are mainly recorded using the 10/20 setup, multiple other datasets use different setups with both fewer, more, and different leads \cite{1300799, BLANKERTZ2007539, HBM:HBM23730, kaya2018large}. This discrepancy in channel setups makes transfer learning across datasets more difficult, since one has to resort to different mapping strategies and possibly discard important information. This issue is especially prevalent with the emergence of large self-supervised models that require large amounts of data for training and in turn should generalize well across a multitude of datasets. Restricting the channel setup in the pretraining or fine-tuning phase therefore limits the usability of these models. 

Encouraged by our previous work~\cite{10285993}, we set out to explore the effect of the pretraining strategy for a channel-agnostic encoder model. To initially assess the performance in a controlled setting, we tested two different pretraining strategies on synthetic data. Here, the most important finding was that failure to match pretraining strategy to the pretraining dataset was detrimental to downstream performance. This indicates that the pretraining strategy should be founded in existing knowledge about the data at hand. 

Moving on to real biomedical time series data, we first assessed the performance on EEG data. Here, our models outperformed the BENDR model and especially, the CRLC pretraining strategy proved beneficial for generalization across data settings. While the model was pretrained on a dataset of only 6 channels, the generalization was shown on two different downstream datasets of 2 and 20 channels, respectively. 

Subsequently the exploration was done for ECG data, and the results were less clear. When looking only at the results of the same model trained with different pretraining strategies, the CRLC pretraining strategy still outperformed the remaining pretraining strategies when combined with the TS2Vec loss. 
The superiority of the CRLC strategy to e.g. the CSC strategy for ECG signals is contrary to what has previously been shown in \cite{kiyasseh2021clocs}. Here, the authors claimed the opposite. Their results were based on the NT-Xent loss, which could hold part of the explanation. Another explanation could be that we do not force the model to align the representations directly between individual leads, but instead between groups of leads. Our strategy would allow for more diversity in the representation of the individual leads. 

Nonetheless, the state-of-the-art reference model outperformed all of the remaining models. This big gap in performance may be caused by the large difference in model size ($5.9\cdot10^4$ vs. $9.0\cdot10^7$ parameters). A potential addition to this may be the fact that all models are pretrained and fine-tuned on datasets that contain the same set of input channels. For ECG data, most public datasets contain a maximum of 12 leads making the mismatch between datasets less of an apparent issue. The smaller mismatch makes it a more viable solution to train a model on the largest number of available leads and zero-pad missing leads in downstream settings. However, recent work also shows that the addition of three leads to the usual 12 contributes to faster and more accurate diagnosis of acute myocardial infarction \cite{vogiatzis2019importance}. These additional leads would be discarded by the reference model and as such, further work on how to improve the performance of channel agnostic models for ECG has the potential to aid in critical situations. 

Finally, we assessed the effect of the aggregation strategy across all datasets by additionally pretraining models with $K=0$ message passing rounds prior to channel aggregation. The results showed that $K=0$ is optimal for the smaller sample sizes, but increasing the number of samples closes the gap and flips the advantage to the more complicated MPNN with $K=3$. These results could be a testiment to the increased number of parameters with the added message passing layers making the models more likely to overfit. The effect of the added message passing rounds is most prominent for the ECG data indicating that the models are less likely to overfit on the ECG data, or simply that the data distribution is more aligned between training, validation and test sets. The results indicate that more research is necessary to determine the optimal aggregation strategies. In connection with this all models are trained using a fully connected graph to encode the spatial information in the signals. Future research should explore more informed graph structures for biosignals, which could additionally aid the issue of overfitting. 

Finally, the MPNN setup used throughout this paper is not restricted to be used with the proposed encoder. Thus, investigating more powerful encoders along with the MPNN could potentially lead to higher performance in both EEG and ECG settings.

\section{Conclusion}
At the dawn of large foundation models for vision and text, research is still necessary to adapt the advances to more complex and scarce data types, such as biomedical time series. Foundation models for multivariate biomedical time series are faced with a number of unique challenges ranging from incorporation of long time dependencies and non-stationarity to model adaptation between varying electrode setups. This work focuses on the latter and specifically on optimal pretraining strategies in this scenario.
To ensure channel-agnosticity, we focus our work on a single-channel encoder followed by a message passing neural network (MPNN) that allows seamless transfer across channel settings. 

Specifically, we propose three different categories of pretraining strategies in the context of contrastive learning. Contrastive random lead coding (CRLC) randomly divides the variables of the multivariate time series into two groups that are then used to form the positive pairs. Contrastive segment coding (CSC) divides windows of time series data into halves along the time axis and uses the two sub-windows as positive pairs. Finally, contrastive augment coding (CAC) applies random augmentations to form the positive pairs. 

CRLC and CSC makes two different assumptions about the inherent nature of the data. Namely, CRLC assumes that the multiple variables share a common source, while CSC assumes stationarity between the two windows. To test the two strategies in scenarios where these assumptions do and do not hold, we created two synthetic datasets obeying either of the two assumptions on which we pretrained models using each strategy. We then fine-tuned the models on two fine-tuning datasets that again obeyed either assumption. Our main finding was that the most important predictor of performance was whether the pretraining strategy matched the pretraining dataset. As such, some knowledge about the data at hand is preferable. 

We then chose two real-life data types on which to test the strategies, namely EEG and ECG. For EEG, channel-agnosticity is especially important. EEG data is typically recorded using a large variety of channel setups. Here, we pretrain a model on a sleep dataset comprising 6 input channels and fine-tune on two different downstream datasets: one containing 2 input channels, both of which are not present in the original 6 (this dataset is referred to as SleepEDFx), and one containing 20 input channels, some of which are among the original 6 (this dataset is referred to as MASS). For both EEG datasets, the CRLC strategy outperforms the remaining two. Additionally, the models achieve higher accuracy than the state-of-the-art reference model (BENDR \cite{BENDR}). The reference model is trained on a fixed set of input channels adhering to the 10/20 montage using masking and reconstruction of tokenized data. These results therefore indicate the continued advantage of contrastive learning, specifically when positive pairs are formed using the CRLC strategy. 

For ECG data, the channel-agnosticity is less important since most ECG data contains a maximum of 12 input channels, the locations of which are typically fixed. For ECG data, the state-of-the-art reference model (W2V+RLM+CMSC \cite{leadagnosticecg}) also takes a fixed set of input channels and is therefore not channel-agnostic. However, the model is trained using random lead masking (RLM), which prepares the model for fewer channels during fine-tuning. Our results showed that this approach was beneficial for ECG, since this model outperformed our remaining pretraining strategies. However, if presented with more channels, this setup would not allow for fine-tuning on the additional inputs. 
When comparing pretraining strategies, the CRLC again outperformed CRLC and CAC. 

Finally, our ablation study on the MPNN indicated that the main benefit is found when fine-tuning on more data. Additionally, the effect of the MPNN was more profound in ECG data. This indicates that the models are more prone to overfitting on EEG data. We therefore believe that future research should investigate other channel aggregation strategies. This could involve using attention layers for combining channels or potentially message passing on spatially informed graphs. 

In conclusion, the CRLC strategy showed promising results when training channel-agnostic models across two data modalities, namely EEG and ECG. 
A potential explanation could be that using groups of channels as positive pairs allows the model to extract the underlying common signal useful for classification. We hope this research will pave the way for future investigations of channel-agnostic self-supervised models for multivariate biosignals.



\section*{References}
\bibliographystyle{IEEEtran}
\bibliography{refs}

\end{document}